\documentclass[aps,prapplied,reprint,groupedaddress]{revtex4-2}
\usepackage{lineno}
\usepackage{graphicx}
\usepackage{subfigure}
\usepackage{float}
\usepackage{dcolumn}
\usepackage{makecell}
\usepackage{bm}
\usepackage{upgreek}
\usepackage[colorlinks=true,linkcolor=blue,citecolor=blue,urlcolor=blue]{hyperref}
\makeatletter

\newcommand{\Rmnum}[1]{\expandafter\@slowromancap\romannumeral #1@}
\makeatother

\begin{document}


\title{Diffuse Laser Cooling Based on the $6\mathrm{P}_{3/2}$ Excited State of Rubidium Atoms via 420~nm Blue Light}


\author{Jia Zhang$^{1}$}
\author{Xun Gao$^{1}$}
\author{Zheng Xiao$^{1}$}
\author{Xiaolei Guan$^{1}$}
\author{Ruihang Chen$^{1}$}
\author{Mengyuan Han$^{1}$}
\author{Tiantian Shi$^{2}$}
\email{tts@pku.edu.cn}
\author{Jingbiao Chen$^{1,3}$}
\affiliation{$^{1}$ State Key Laboratory of Advanced Optical Communication Systems and Networks, Institute of Quantum Electronics, School of Electronics, Peking University, Beijing 100871, China}
\affiliation{$^{2}$ National Key Laboratory of Advanced Micro and Nano Manufacture Technology, School of Integrated Circuits, Peking University, Beijing 100871, China}
\affiliation{$^{3}$ Hefei National Laboratory, Hefei 230088, China}

\date{\today}

\begin{abstract}
To date, the laser cooling of rubidium atoms has inevitably relied on 780~nm cooling light corresponding to the first excited state $5\mathrm{P}_{3/2}$. Surprisingly, we demonstrate laser cooling directly utilizing 420~nm blue light for active optical clock, which corresponds to the high excited state $6\mathrm{P}_{3/2}$ of Rb atom. Experimentally, we successfully apply the 420~nm diffuse laser cooling technique to prepare a cold $^{87}\mathrm{Rb}$ atomic cloud with a length of up to one meter, and measure the cold-atom absorption spectroscopy. The cold atom number is approximately $4.4\times10^{7}$. We systematically compare the cooling effects of 420~nm and 780~nm diffuse laser cooling, and verify the feasibility of blue light cooling using high excited state. This work directly employs blue light to cool and manipulate ground-state Rb atoms to the 6P excited state, providing a new and efficient approach for the cold-atom active optical clock. It is also expected to open up research directions and application prospects in frontier fields such as Rydberg atoms, ultracold quantum gases Bose-Einstein condensation, quantum information, and so on.
\end{abstract}

\maketitle


\section{Introduction}
The disorder induced by the atomic thermal motion has long been one of the major obstacles hindering in-depth exploration of internal quantum states~\cite{ciurylo2004photoassociation} and their interactions, precise measurement of fundamental physical constants~\cite{hart2007quantum}, and even the realization of quantum precision manipulation~\cite{shaw2025erasure,wineland1998quantum}. Slowing down and cooling atoms to extremely low temperatures to render them nearly ``stationary" is the key to overcoming these obstacles and opening up new frontiers in modern atomic and molecular physics as well as quantum technology. Laser cooling technology achieves efficient deceleration and cooling of free atoms through momentum exchange between photons and atoms~\cite{chu1985three,raab1987trapping,spencer1959low,metcalf1999laser}, and its emergence and development undoubtedly represent a revolutionary breakthrough. Precise manipulation of atomic motion to near-stationary ultracold states serves as a core methodology in modern atomic and molecular physics~\cite{rosenberg2022observation,weber2003bose}, quantum precision measurement~\cite{zheng2022differential,li2025realization,aeppli2024clock,zhang2023ultrahigh}, and quantum information science~\cite{osborne2025large,halimeh2025cold}.

Among numerous atomic systems, rubidium atoms have become one of the most widely used platforms in cold atom research, owing to their mature laser technology (D2 line $5\mathrm{S}_{1/2}$-$5\mathrm{P}_{3/2}$ with a wavelength of 780~nm~\cite{heavens1961radiative}) and favorable energy level characteristics (electric dipole-allowed strong transitions and closed cycle transitions). Cooling technologies based on 780~nm lasers, such as Doppler cooling~\cite{gerz1993temperature,rosi2018lambda}, polarization gradient cooling~\cite{liang2024multi}, Raman sideband cooling~\cite{hu2017creation,kaufman2012cooling,thompson2013coherence} and magneto-optical trap (MOT)~\cite{goldwin2002two,jarvis2018blue}, have reached a high level of maturity. They can efficiently cool and trap large numbers of rubidium atoms to the Doppler cooling temperature limit or even sub-Doppler temperatures, laying the foundation for numerous cutting-edge research fields including Bose-Einstein condensation (BEC)~\cite{hu2017creation}, atomic clocks~\cite{bohnet2012steady}, and atomic interferometers~\cite{xue2015continuous}. In relevant studies on laser cooling of rubidium atoms, most systems employ cooling and repumping light addressing different hyperfine sublevels within the same set of fine-structure energy levels. 
However, the frequencies of the pumping and cooling light predominantly concentrate on red transitions near 780~nm and 795~nm in rubidium, while research on blue-light cooling corresponding to the 420~nm and 421~nm transitions of the high excited state remains relatively scarce.

Conventional 780~nm cooling schemes possess a relatively large natural linewidth ($\Gamma_{780}/2\pi=6.06$~MHz~\cite{heavens1961radiative,gutterres2002determination}). While this facilitates high photon scattering rates for rapid initial cooling, it also imposes a relatively high Doppler cooling temperature limit. Furthermore, the comparatively short lifetime of the $5\mathrm{P}_{3/2}$ level imposes constraints on further advancing coherent manipulation. Current research utilizing the $6\mathrm{P}_{3/2}$ excited state of rubidium atoms as the cooling level consistently integrates both 780~nm and 420~nm cooling~\cite{das2023narrow,das2024continuous}. The first approach involves preliminary cooling with 780~nm lasers, followed by secondary cooling using 420~nm blue light~\cite{das2023narrow}. In essence, it utilizes the broad linewidth of 780~nm ($\sim$6.06~MHz) to achieve efficient trapping and initial deceleration, and then leverages the narrow linewidth of 420~nm blue light ($\Gamma_{420}/2\pi=1.42$~MHz~\cite{marek1980radiative,safronova2011critically} to break through the temperature limit of 780~nm laser cooling. This process necessitates laser source switching and critically depends on the preliminary cooling stage. Another approach is the dual-light cooperative cooling scheme~\cite{das2024continuous}: both 780~nm and 420~nm lasers are turned on simultaneously, but the 780~nm light still serves as the dominant cooling light, with the 420~nm light playing an auxiliary role. This configuration retains a fundamental reliance on the 780~nm light source and introduces complexities associated with spectral interference and power-matching requirements.

In this work, we completely abandon the 780~nm cooling optical path and achieve cooling of $^{87}\mathrm{Rb}$ atoms solely through 420~nm cooling laser. Combined with diffuse laser cooling, we directly utilize a high-power 420~nm blue laser as the cooling light to cool $^{87}\mathrm{Rb}$ atoms, realizing a cold $^{87}\mathrm{Rb}$ atomic cloud with a meter-scale length. Experimentally, we utilize a 780~nm laser as the probe light and observe the absorption spectral line of cold $^{87}\mathrm{Rb}$ atoms. In addition, we simultaneously conduct research on 780~nm diffuse laser cooling of $^{87}\mathrm{Rb}$ atoms and performe a comparison with the direct 420~nm laser cooling. This work demonstrates that rubidium atoms can achieve cooling through the high-excited-state pathway ($5\mathrm{S}_{1/2}$-$6\mathrm{P}_{3/2}$), challenging the traditional perception that ``low-excited-state transitions are an essential pathway for cooling". Through this work, we aim to provide practical experience for overcoming the technical bottlenecks of blue light cooling, verify its unique advantages over traditional cooling methods, and utilize blue light to cool and manipulate ground-state atoms to the 6P state. This provides simultaneous cooling and pumping for the realization of cold-atom active optical clock~\cite{chang2019stabilizing,zhang2024extremely}, and  offers a new idea for the efficient preparation and coherent manipulation of Rydberg atoms (promising candidates for quantum computation and quantum simulation)~\cite{ryabtsev2016spectroscopy}, 420~nm qoptical frequency standard~\cite{zhang2017420,zhang2017compact}, atom interferometry with rubidium blue transitions~\cite{salvi2023atom}, and so on.

\section{THEORETICAL ANALYSIS}
Narrow-linewidth laser cooling has been a key technology for realizing cold atoms of alkaline-earth elements (such as calcium~\cite{curtis2003quenched,sterr2003prospects}, strontium~\cite{loftus2004narrow}, and ytterbium~\cite{guttridge2016direct}) and rare-earth elements (such as erbium~\cite{frisch2012narrow}, dysprosium~\cite{maier2014narrow}, and thulium~\cite{sukachev2014secondary}). For hydrogen-like alkali atoms, the feasibility of narrow-linewidth cooling has been demonstrated using $\mathrm{nS_{1/2}}\rightarrow\mathrm{(n+1)P_{3/2}}$ transitions in lithium (323~nm $\mathrm{2S_{1/2}}\rightarrow\mathrm{3P_{3/2}}$)~\cite{duarte2011all,sebastian2014two}, potassium (405~nm $\mathrm{4S_{1/2}}\rightarrow\mathrm{5P_{3/2}}$)~\cite{mckay2011low}, and rubidium (420~nm $\mathrm{5S_{1/2}}\rightarrow\mathrm{6P_{3/2}}$)~\cite{das2023narrow,das2024continuous}. 
Due to the narrow natural linewidth of the 420~nm $\mathrm{5S_{1/2}}\rightarrow\mathrm{6P_{3/2}}$ transition in rubidium atoms, the corresponding Doppler cooling temperature limit $T_{\mathrm{D}}=\frac{\hbar\Gamma}{2k_{\mathrm{B}}}$~\cite{lett1989optical} ($\hbar$ is the reduced Planck constant, $\Gamma$ represents the natural linewidth, $k_{\mathrm{B}}$ is the Boltzmann constant) is lower than that corresponding to the 780~nm transition. For an ideal gas of atoms in thermal equilibrium (neglecting interactions), their velocities obey the Maxwell-Boltzmann velocity distribution law.  The most probable velocities ($v_{\mathrm{p}}=\sqrt{\frac{2k_{\mathrm{B}}T}{m}}$,where $k_\mathrm{B}$ is the Boltzmann constant, $T$ is the atomic temperature, and $m$ is the atomic mass) corresponding to the Doppler cooling temperature limits for the two wavelengths of 420~nm and 780~nm are 0.081~m/s and 0.167~m/s, respectively. A comparison between 780~nm Doppler cooling and 420~nm Doppler cooling is presented in Table~\ref{table1}.

\begin{table}[thp]
	\caption{\label{table1}%
		Comparison between 780~nm Doppler cooling and 420~nm Doppler cooling.}
	\begin{ruledtabular}
		\begin{tabular}{lllll}
			$\lambda$ (nm)&\begin{tabular}[c]{@{}c@{}}Corresponding\\ energy levels\end{tabular}&\begin{tabular}[c]{@{}c@{}}$\Gamma$ (MHz)\end{tabular}&  $T_{\mathrm{D}}$ ($\mu\mathrm{K}$)&$v_{\mathrm{p}}$ (m/s)\\
			\colrule\rule{0pt}{1.2em}%
			780&$\mathrm{5S_{1/2}}\rightarrow\mathrm{5P_{3/2}}$&6.06~\cite{heavens1961radiative,gutterres2002determination}&146&0.167\\
			420&$\mathrm{5S_{1/2}}\rightarrow\mathrm{6P_{3/2}}$&1.42~\cite{marek1980radiative,safronova2011critically}&34&0.081\\
		\end{tabular}
	\end{ruledtabular}
\end{table}

\begin{figure*}
	\includegraphics[width=\linewidth]{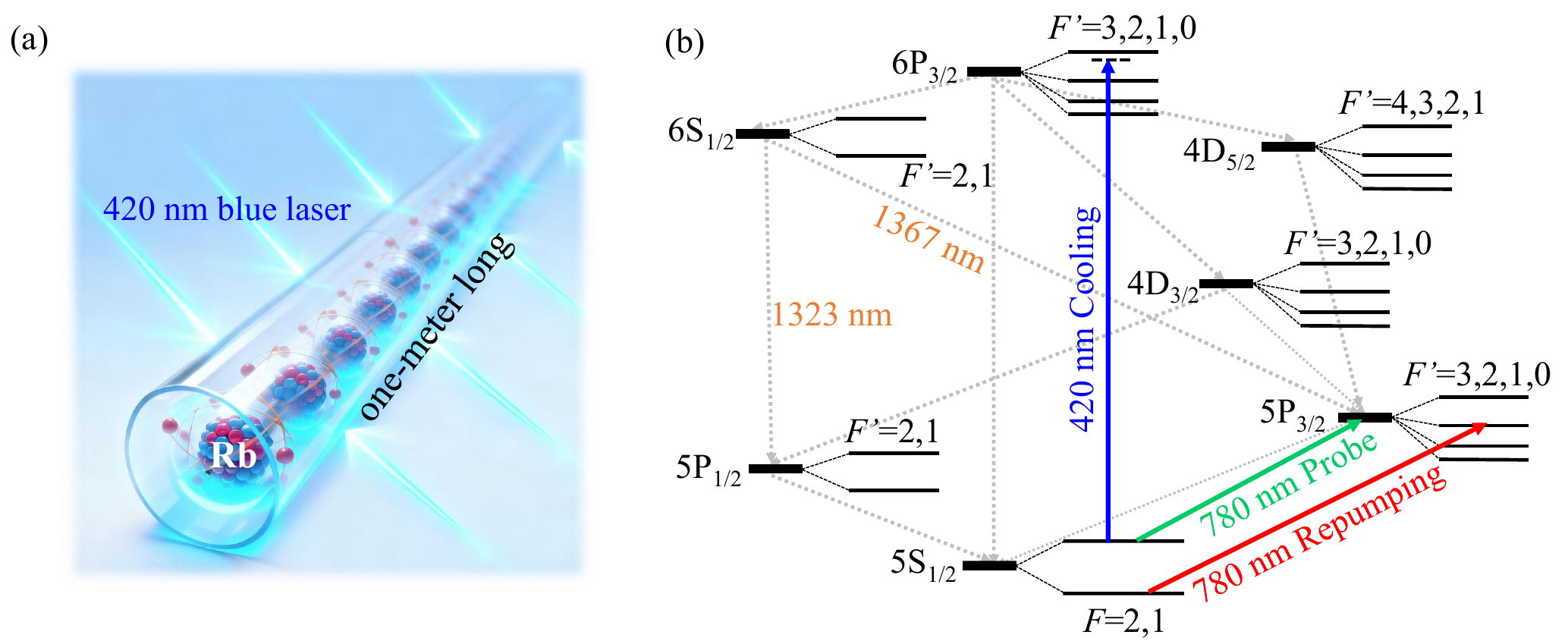}
	\caption{\label{figure-1}
		(a) Diagram of 420~nm blue light diffuse laser cooling, which produced a cold rubidium atom cloud up to one meter in length. (b) Relevant energy levels involved in 420~nm blue light cooling. The cooling light has a certain red detuning relative to the 420~nm $\mathrm{5S_{1/2}}(F=2)\rightarrow\mathrm{6P_{3/2}}(F^{\prime}=3)$ transition. The repumping light corresponds to the 780~nm $\mathrm{5S_{1/2}}(F=1)\rightarrow\mathrm{5P_{3/2}}(F^{\prime}=2)$ transition and the probe light corresponds to the 780~nm $\mathrm{5S_{1/2}}(F=2)\rightarrow\mathrm{5P_{3/2}}$ transition. Simultaneously, the 420~nm laser cooling light also functions as a pumping light, creating a population inversion between the $\mathrm{6S_{1/2}}$ and $\mathrm{5P_{3/2}}$, $\mathrm{5P_{1/2}}$ energy levels, which facilitates the realization of lasing at 1367~nm and 1323~nm for the cold-atom active optical clock. 
	}
\end{figure*}

Concurrently, 420~nm blue-light cooling faces significant challenges: 1) Laser System Complexity: Generating high-power, narrow-linewidth, frequency-stablized 420~nm blue lasers directly is less mature and cost-effective compared to 780~nm lasers. It typically requires frequency-doubling techniques (e.g., doubling 840~nm laser light to 420~nm), which increases system complexity and introduces power loss. 2) Scattering Rate and Trap Depth: Due to the longer excited-state lifetime (narrower linewidth), the photon scattering rate for the 420~nm transition is lower than that of the 780~nm transition. This adversely affects the trapping force and velocity capture range, necessitating increased cooling light power. 3) Branching Ratios: When atoms decay from the $\mathrm{6P_{3/2}}$ state, only a fraction return directly to the ground state $\mathrm{5S_{1/2}}$; another fraction cascades into intermediate states (e.g., $\mathrm{4D_{3/2}}$, $\mathrm{4D_{5/2}}$, $\mathrm{6S_{1/2}}$), as shown in Fig.~\ref{figure-1}. Therefore, investigating the decay branching ratios from the 6P state to these intermediate states is crucial for designing an effective repumping scheme to prevent atom loss due to optical pumping.

\begin{figure}
	\includegraphics[width=0.9\linewidth]{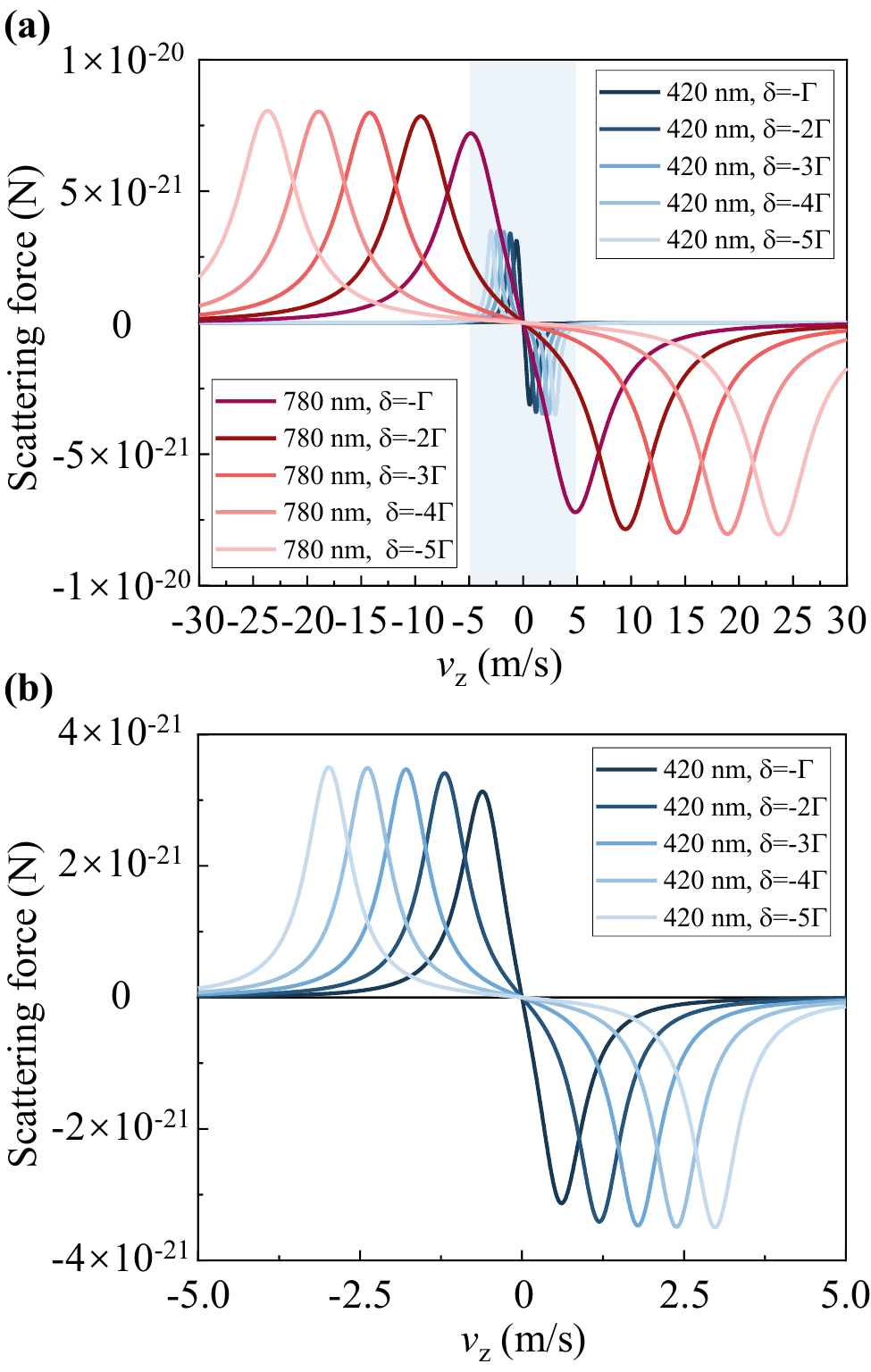}
	\caption{\label{figure-2}
		In the calculation, the saturation coefficient $s_{0}$ is set to 1, and the one-dimensional optical molasses model is used to calculate the variation of the scattering force with the velocity of $^{87}\mathrm{Rb}$ atoms under different detuning $\delta$ conditions.  (a) 780~nm laser as cooling light; (b) 420~nm laser as cooling light. When both the 780~nm cooling light and the 420~nm cooling light are at saturation intensity, with the same detuning, the velocity of the atomic group acted on by the 780~nm cooling light is greater than that of the 420~nm cooling light.
	}
\end{figure}

In the experiment, we respectively employ 780~nm cooling light and 420~nm cooling light to cool $^{87}\mathrm{Rb}$ atoms, and theoretically investigate the characteristics of scattering force and detuning optimization of $^{87}\mathrm{Rb}$ atoms in these two cooling systems. To find the optimal experimental parameters, a relational model between atomic velocity and scattering force is established under the condition of fixed detuning. Under fixed red detunings, the scattering force generated by the 780~nm cooling light and 420~nm cooling light exhibits different characteristics in velocity space.
\begin{equation}
	F_{\mathrm{scatt}}=\frac{\hbar k\Gamma}{2}\left[\frac{s_{0}}{1+s_{0}+\frac{2(\delta-kv_{\mathrm{z}})}{\Gamma}}-\frac{s_{0}}{1+s_{0}+\frac{2(\delta+kv_{\mathrm{z}})}{\Gamma}}\right]
\end{equation}
Among them, $\hbar$ is the reduced Planck constant, $k$ is the wave vector, $s_{0}$ is the saturation coefficient, $\delta$ is the detuning, and $v_{\mathrm{z}}$ is the atomic velocity. For simplicity, consider the total scattering force of a pair of cooling lights in relative directions. As shown in Fig.~\ref{figure-2}(a), the 780~nm cooling light shows a more gradual rising trend in the low-velocity region, while the 420~nm cooling light exhibits a steeper growth slope. This indicates that there are different trapping abilities in capturing low-velocity atoms between the two cooling lights. When the intensity of both cooling lasers is set to their respective saturation intensities, and the detuning of the 420~nm and 780~nm cooling lasers is equal to -n times their natural linewidths, the atomic velocity corresponding to the maximum scattering force of the 420~nm cooling laser is significantly lower than that of the 780~nm cooling laser. 

Fig.~\ref{figure-2}(b) illustrates the relationship between the scattering force generated by the 420~nm cooling laser and atomic velocity. Within a certain range, as the detuning increases, the speed range of the capture atoms will also increase. When the 420~nm cooling laser is detuned by $-5\Gamma$, the atomic velocity at the maximum scattering force is only $\pm$2.98~m/s. Near this velocity, the photon absorption-reemission cycle rate (i.e., scattering force) reaches its peak, subjecting low-velocity atomic ensembles in this range to the strongest deceleration and trapping effects. Atoms far from this velocity (particularly high-velocity atoms) experience weak forces and are thus difficult to trap effectively. Consequently, under the 420~nm cooling laser field, the decelerated atoms are predominantly low-velocity particles. Moreover, as can be seen from the Fig.~\ref{figure-2}(a), the atomic velocity range corresponding to the scattering force generated by the 420~nm cooling light is narrower than that of the 780~nm cooling light, which corresponds to a smaller number of cooled atoms.

For different red detuning conditions of the cooling light, the optimal detuning parameters are sought by solving the maximum value of the velocity-dependent damping coefficient $\beta=-\frac{\partial F_{\mathrm{scatt}}}{\partial v}|_{v=0}$. The theoretical model is based on the Doppler cooling limit, where the damping coefficient reaches its optimal value $\beta_{\mathrm{opt}}=\hbar k^{2}$ when the detuning $\delta=-\frac{\Gamma}{2}$. Therefore, the optimal red detunings for the 780~nm and 420~nm cooling lights are 3~MHz and 0.7~MHz, respectively. However, subsequent experimental measurements reveal that the optimal detuning for the 780~nm system is $\delta_{\mathrm{opt, 780}}=-15$~MHz (approximately 2.5 times its natural linewidth $\Gamma_{780}/2\pi=6.06$~MHz), while the 420~nm system requires $\delta_{\mathrm{opt, 420}}=-8.0$~MHz, which is related to its narrower linewidth $\Gamma_{420}/2\pi=1.42$~MHz. The theoretical value $\delta=-\frac{\Gamma}{2}$ is derived for an ideal two-level atom under the weak light intensity limit, considering only the Doppler cooling mechanism, which yields the maximum damping coefficient (fastest cooling rate). In real systems, particularly alkali metal atoms, the effective scattering rate is reduced due to the multi-level structure (optical pumping), the maximum radiation pressure shifts to larger detunings under high light intensity (saturation effect), and other practical constraints (magnetic fields, trapping range, collisions, etc.) come into play. Therefore, in experiments, to achieve the lowest temperature, sufficient trapping efficiency, and stable system operation, the red detuning of the cooling light needs to be optimized over a broader range. This optimal value aims to maximize the overall cooling performance (final temperature, trapped atom number, stability) rather than just the theoretical maximum instantaneous cooling rate.

\section{EXPERIMENT} 
\begin{figure*}
	\includegraphics[width=\linewidth]{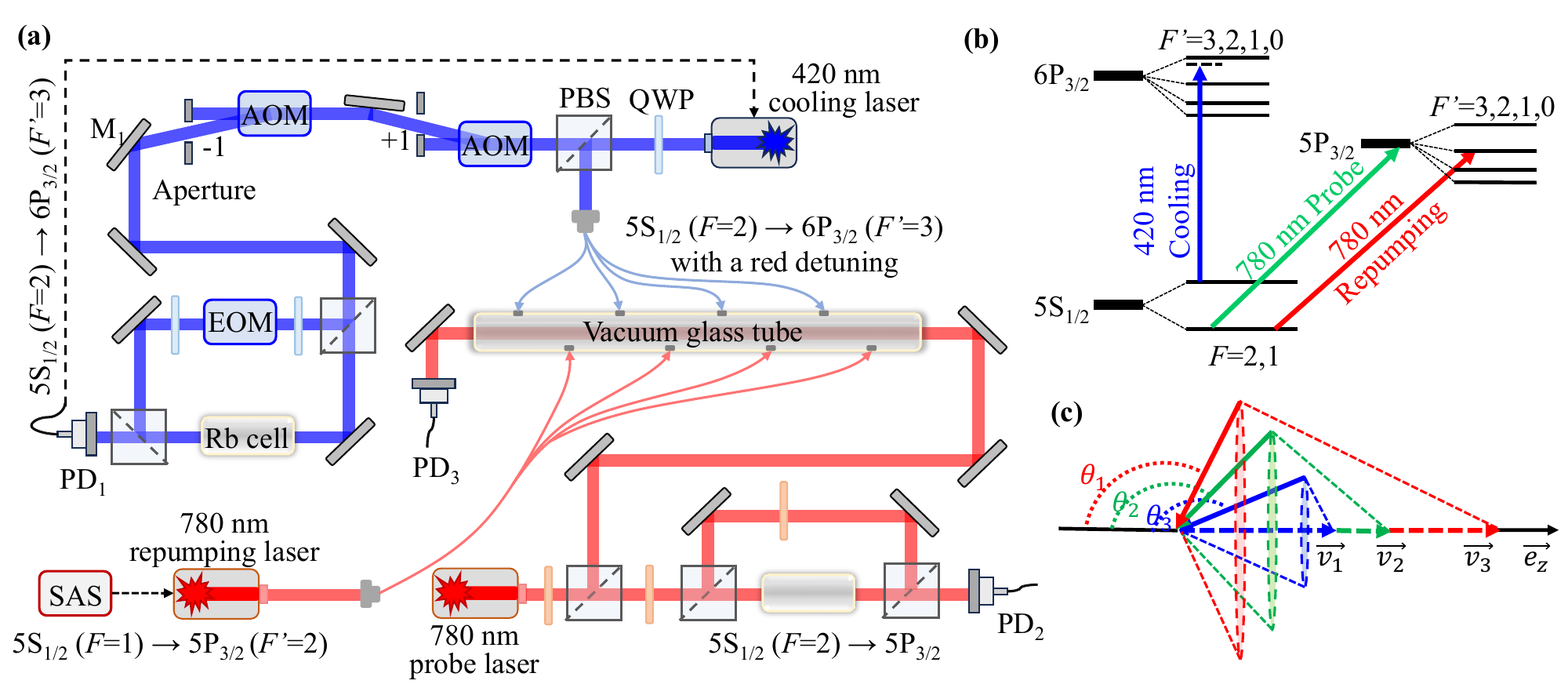}
	\caption{\label{figure-4}
		Experimental system. (a) The 420~nm laser undergoes frequency shifting before frequency locking, resulting in a 420~nm cooling light with a specific red detuning relative to the $\mathrm{5S_{1/2}}(F=2)\rightarrow\mathrm{6P_{3/2}}(F^{\prime}=3)$ transition. The 780~nm repumping light is locked to the $\mathrm{5S_{1/2}}(F=1)\rightarrow\mathrm{5P_{3/2}}(F^{\prime}=2)$ transition, while the probe light corresponds to the 780~nm $\mathrm{5S_{1/2}}(F=2)\rightarrow\mathrm{5P_{3/2}}$ transition. (b) Relevant energy levels. (c) Schematic illustration of the principle of diffuse reflection cooling.
	}
\end{figure*}

Due to the narrower natural linewidth of the 420~nm transition ($\sim1.42$~MHz), its theoretical maximum scattering rate is lower than that of the 780~nm transition, resulting in insufficient trapping force in traditional schemes to directly cool thermal atoms. Therefore, in this experiment, we specifically conduct a preliminary investigation on a watt-level high-power 420~nm laser. Using modulation transfer spectroscopy, we achieve frequency locking of the 420~nm laser with a 3~W power output~\cite{zhang2024power}. After performing appropriate frequency shift, the laser is used directly as cooling light. The experimental system, as shown in Fig.~\ref{figure-4}(a), includes 420~nm cooling light, 780~nm repumping light, 780~nm probe light, a rubidium atomic vacuum cell, and a vacuum maintenance module.

Due to the longer lifetime of the $6\mathrm{P}_{3/2}$ excited state of rubidium atoms (resulting in a narrow linewidth), the photon scattering rate of the 420~nm transition is lower than that of the 780~nm transition. To enhance the cooling effect, a higher power of the 420~nm cooling light is required. In the experiment, we design and constructed a stable high-power 420~nm blue laser system based on frequency doubling technology. 1532~nm and 1862~nm lasers serve as seed lasers. Via sum frequency generation, they produce an 840~nm laser, which is subsequently converted into the target 420~nm laser through frequency doubling. The output power of this 420~nm laser can reach 3~W. In the experiment, appropriate frequency shifting of the 420~nm laser is necessary to introduce a certain red detuning relative to the $\mathrm{5S_{1/2}}(F=2)\rightarrow\mathrm{6P_{3/2}}(F^{\prime}=3)$ transition of $^{87}\mathrm{Rb}$ atoms. The relevant partial energy levels involved are shown in Fig.~\ref{figure-4}(b).

Generally, the cooling laser is first frequency-locked and then shifted using an acousto-optic modulator (AOM). However, AOMs in the 420~nm blue wavelength range exhibit low diffraction efficiency, leading to significant power loss when using the conventional frequency-locking-then-shifting approach. Therefore, we adopt a method of frequency shifting before frequency locking. The natural linewidth of the 420~nm transition is 1.42~MHz, and the final red detuning is typically on the order of MHz. Since commercial AOMs usually have center frequencies in the tens of MHz, two AOMs are employed to shift the frequency using their +1st and -1st order diffracted beams, respectively. This configuration results in a blue detuning of several MHz relative to the $\mathrm{5S_{1/2}}(F=2)\rightarrow\mathrm{6P_{3/2}}(F^{\prime}=3)$ transition of $^{87}\mathrm{Rb}$. Subsequently, the frequency-shifted 420~nm laser is locked to the $\mathrm{5S_{1/2}}(F=2)\rightarrow\mathrm{6P_{3/2}}(F^{\prime}=3)$ transition line using modulation transfer spectroscopy, yielding a 420~nm cooling light with a red detuning of several MHz relative to this transition.

Experimentally, we use a rubidium source of natural isotopic composition, consisting of approximately 72\% $^{85}\mathrm{Rb}$ and 28\% $^{87}\mathrm{Rb}$. Rubidium atoms pumped to the $\mathrm{6P_{3/2}}$ state by the 420~nm cooling light may decay to energy levels such as $\mathrm{4D_{3/2}}$, $\mathrm{4D_{5/2}}$, and $\mathrm{6S_{1/2}}$, causing them to escape from the cooling cycle. Therefore, a 780~nm laser is locked to the $^{87}\mathrm{Rb}$ $\mathrm{5S_{1/2}}(F=1)\rightarrow\mathrm{5P_{3/2}}(F^{\prime}=2)$ transition line via saturated absorption spectroscopy to serve as repumping light. This returns the atoms to the $\mathrm{5S_{1/2}}(F=2)$ state, allowing them to re-enter the 420~nm blue-light cooling process. 
The 420~nm cooling light and 780~nm repumping light are coupled into multimode fibers and injected into a one-meter-long vacuum atomic cell. The cylindrical cell, fabricated from glass, has a length of 1 meter and a diameter of 2 centimeters. The vacuum environment is maintained by an ion pump with a pumping speed of 45~L/s, achieving a vacuum level on the order of $10^{-6}$ Pa. The outer wall of the vacuum cell is exposed to the laboratory environment, with the ambient temperature maintained at 22~$^{\circ}$C.

In this work, the diffuse laser cooling technique~\cite{ketterle1992slowing,batelaan1994slowing,wan2022quasi,wang2022realizing} is employed to cool rubidium atoms, and its principle is similar to that of molasses. This approach eliminates the need for an optical trap structure, thereby producing a large-volume cold-atom cloud without applying magnetic fields. The diffuse light field exhibits identical light intensity and frequency characteristics in all directions. This ensures that regardless of the direction a rubidium atom moves, it can absorb photons whose frequency matches its velocity (due to the Doppler effect). Upon absorption, the atom loses momentum through spontaneous emission, thereby achieving cooling. As shown in the schematic diagram in Fig.~\ref{figure-4}(c), atoms have different velocities ($\vec{v_{1}}$, $\vec{v_{2}}$, and $\vec{v_{3}}$) in the direction of $\vec{v_{\mathrm{z}}}$. The cooling light propagates at an angle $\theta$ to the atomic velocity vector. The frequency $\omega$ of the cooling light is detuned from the atomic transition resonance frequency $\omega_{\mathrm{a}}$ by an amount $\delta$. Absorption occurs when the atom's velocity satisfies the condition $\delta=\omega\cdot v \mathrm{cos}\theta/c$, where $c$ is the speed of light. In the experiment, the detuning $\delta$ is fixed. Consequently, the component of the atomic velocity along the propagation direction of the light, $v \mathrm{cos}\theta$, is also fixed. Atoms automatically absorb cooling light at the appropriate angle $\theta$ to compensate for the Doppler shift, enabling cooling. Theoretically, atoms within this isotropic light field can always absorb specific Doppler-shifted light to be cooled. Therefore, isotropic cooling via this diffuse light field is more efficient than traditional molasses cooling generated by three orthogonal pairs of counter-propagating laser beams.

To generate a diffuse laser cooling light field (also referred to as an isotropic light field), a barium sulfate ($\mathrm{BaSO_{4}}$) diffuse reflection coating is uniformly applied to the outer wall of the vacuum atomic cell. The 420~nm cooling light undergoes diffuse reflection via the high-reflectivity surface of this material, thereby forming a uniformly distributed isotropic cooling light field. The barium sulfate coating has a reflectivity of approximately 98\% for 780~nm laser and approximately 95\% for 420~nm laser. After injecting the cooling light and repumping light into the vacuum atomic cell, the diffuse-coated vacuum cell appears blue. When rubidium atoms move in the light field, due to the Doppler effect, the light frequency perceived by the atoms varies with their direction of motion. Regardless of the direction in which the rubidium atoms move, cooling can always be achieved. Owing to the isotropy of the light field, the diffuse reflection cooling process is more uniform, which can effectively reduce the atomic velocity distribution. In addition, diffuse laser cooling does not require complex laser beam collimation and counter-propagation, making the realization of isotropic light cooling easier and simpler. In the experiment, a homemade 780 nm interference filter configuration external cavity diode laser is used as the probe light~\cite{guan2025780}. This probe beam is directed through the one-meter-long vacuum atomic cell, enabling the detection of the cold $^{87}\mathrm{Rb}$ atoms.

\section{RESULTS} 
\subsection{Absorption spectroscopy of cold $^{87}\mathrm{Rb}$ atoms}
\begin{figure}
	\includegraphics[width=\linewidth]{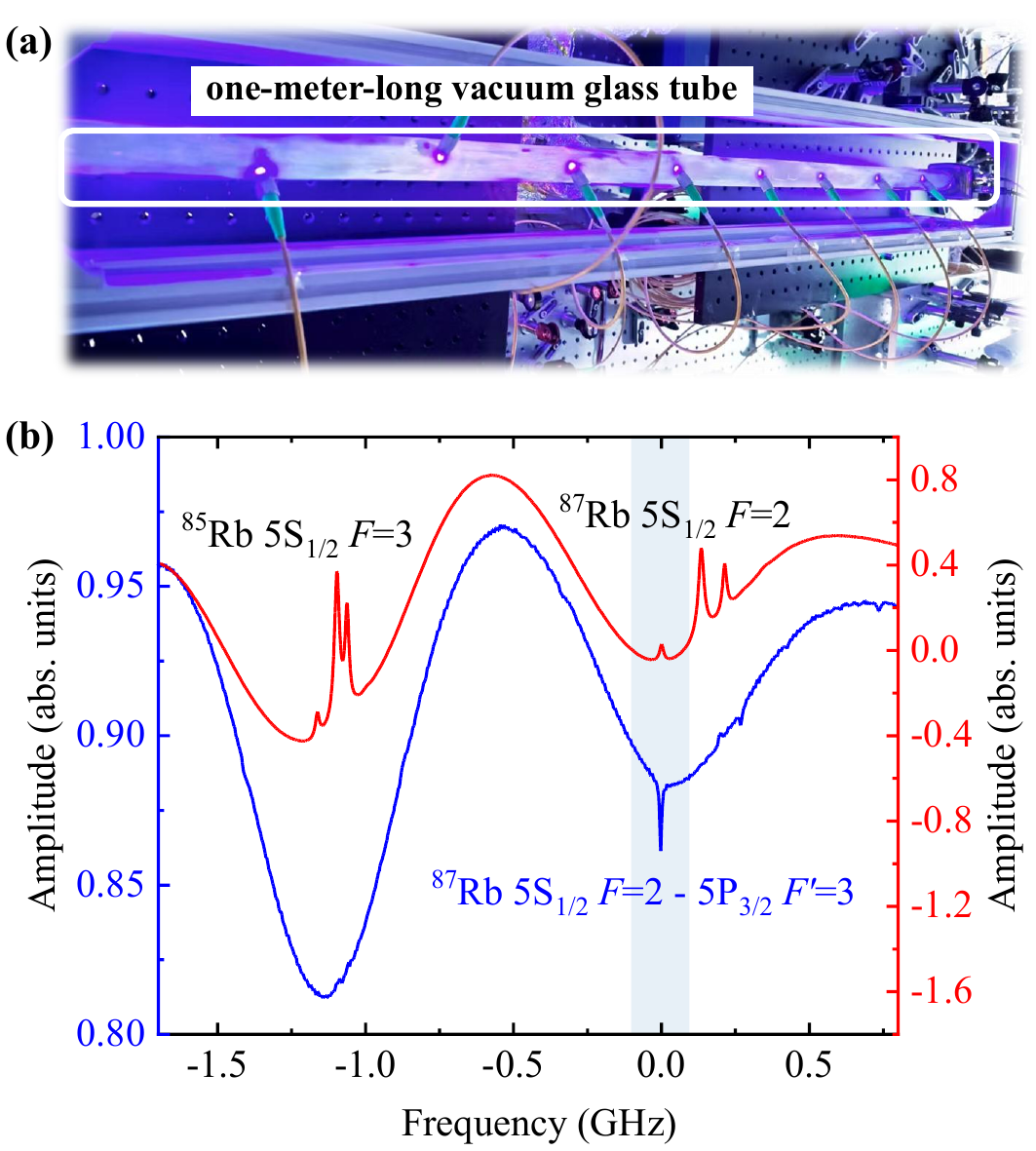}
	\caption{\label{figure-5}
		(a) Photograph of the vacuum atomic cell. Under the cooling effect of the diffuse light field, a cold $^{87}\mathrm{Rb}$ atomic cloud with a length of up to one meter is formed.  
		(b) Absorption spectroscopy of cold $^{87}\mathrm{Rb}$ atoms for the 780~nm probe light (blue curve). The red curve serves as the reference rubidium saturated absorption spectroscopy. Obvious cold atomic absorption can be observed at the position corresponding to the $^{87}\mathrm{Rb}$  $\mathrm{5S_{1/2}}(F=2)\rightarrow\mathrm{5P_{3/2}}(F^{\prime}=3)$ transition.
		}
\end{figure}

The frequency of the 420~nm cooling light is red-detuned by approximately 8~MHz relative to the $^{87}\mathrm{Rb}$ $\mathrm{5S_{1/2}}(F=2)\rightarrow\mathrm{6P_{3/2}}(F^{\prime}=3)$ transition. Meanwhile, a 30~mW 780~nm repumping laser is applied, whose frequency is precisely tuned to the $\mathrm{5S_{1/2}}(F=1)\rightarrow\mathrm{5P_{3/2}}(F^{\prime}=2)$ transition. Under the action of the diffuse light field, $^{87}\mathrm{Rb}$ atoms are effectively decelerated and cooled, spatially forming a cold $^{87}\mathrm{Rb}$ atomic cloud with a length of up to one meter, as shown in the vacuum atomic cell in Fig.~\ref{figure-5}(a). To characterize the internal energy states and transition properties of the cooled atoms, a weak 780~nm probe light with microwatt-level power is applied axially. The frequency of the probe light is precisely scanned near the rubidium atom D2 line ($\mathrm{5S_{1/2}}\rightarrow\mathrm{5P_{3/2}}$) with a range covering several hundred MHz. When the scanned frequency is resonant with the allowed transition from the atomic ground state ($F=2$) to the excited state ($F^{\prime}$), the cold atoms absorb photons, and the intensity of the probe light attenuates. By using a photodetector to record the variation of transmitted light intensity with the probe light frequency, the cold atomic absorption spectroscopy can be obtained.

In the cold $^{87}\mathrm{Rb}$ atom system, three absorption peaks can be observed when probing with a frequency-scanned 780~nm laser. This phenomenon fundamentally results from the combined effects of the quantum selection rules for transitions between hyperfine energy levels and the extremely narrow velocity distribution of the cold atoms. For the cooled $^{87}\mathrm{Rb}$ atoms, during the cooling process, atoms are selectively prepared in the ground state $F=2$ sublevel via optical pumping. At this point, the probe light can only excite transitions that satisfy the angular momentum selection rule $\delta F=0, \pm1$. These correspond to the three allowed transition channels: $F=2 \rightarrow F^{\prime}=1$, $F=2 \rightarrow F^{\prime}=2$, and $F=2 \rightarrow F^{\prime}=3$. Since the thermal motion velocity of cold atoms is suppressed to the level of tens of $\mu\mathrm{{K}}$, their Doppler broadening ($\delta v=\sqrt{\frac{k_{\mathrm{B}}T/m}{\lambda}}$) is reduced to the kHz level. This is significantly narrower than the hyperfine level spacings (approximately tens to hundreds of MHz). Consequently, adjacent transition peaks, which would be obscured by Doppler broadening in a thermal atomic vapor, become fully resolved, forming three independent and distinguishable Lorentzian absorption peaks. However, for the 420~nm laser cooling in this experiment, the cooling efficiency is inferior to that of traditional 780~nm laser cooling, resulting in a smaller number of cold atoms. Thus, obvious cold atomic absorption is only observed at the position corresponding to the $^{87}\mathrm{Rb}$ $\mathrm{5S_{1/2}}(F=2)\rightarrow\mathrm{5P_{3/2}}(F^{\prime}=3)$ transition, as shown by the blue curve in Fig.~\ref{figure-5}(b). It can be seen that on the background of Doppler absorption of thermal atoms, there exists an obvious and sharp absorption dip of cold atoms.

Due to the significant suppression of atomic thermal motion, the Doppler broadening effect is essentially eliminated. Consequently, the spectral linewidth primarily reflects the natural linewidth of the excited state and possible residual broadening mechanisms (such as collisions). To precisely calibrate the probe light frequency and confirm spectral line, the saturated absorption spectroscopy of rubidium atoms is simultaneously recorded in the experiment as a frequency reference. Utilizing the nonlinear saturation effect to eliminate the Doppler background, this reference spectroscopy clearly reveals the accurate central frequencies of the hyperfine transitions: the $^{87}\mathrm{Rb}$ $\mathrm{5S_{1/2}}(F=2)\rightarrow\mathrm{5P_{3/2}}$ and the $^{85}\mathrm{Rb}$ $^{87}\mathrm{Rb}$ $\mathrm{5S_{1/2}}(F=3)\rightarrow\mathrm{5P_{3/2}}$ transitions. It provides the reference for the accurate identification of the cold atom spectral lines and the linewidth measurement.

\begin{figure}
	\includegraphics[width=\linewidth]{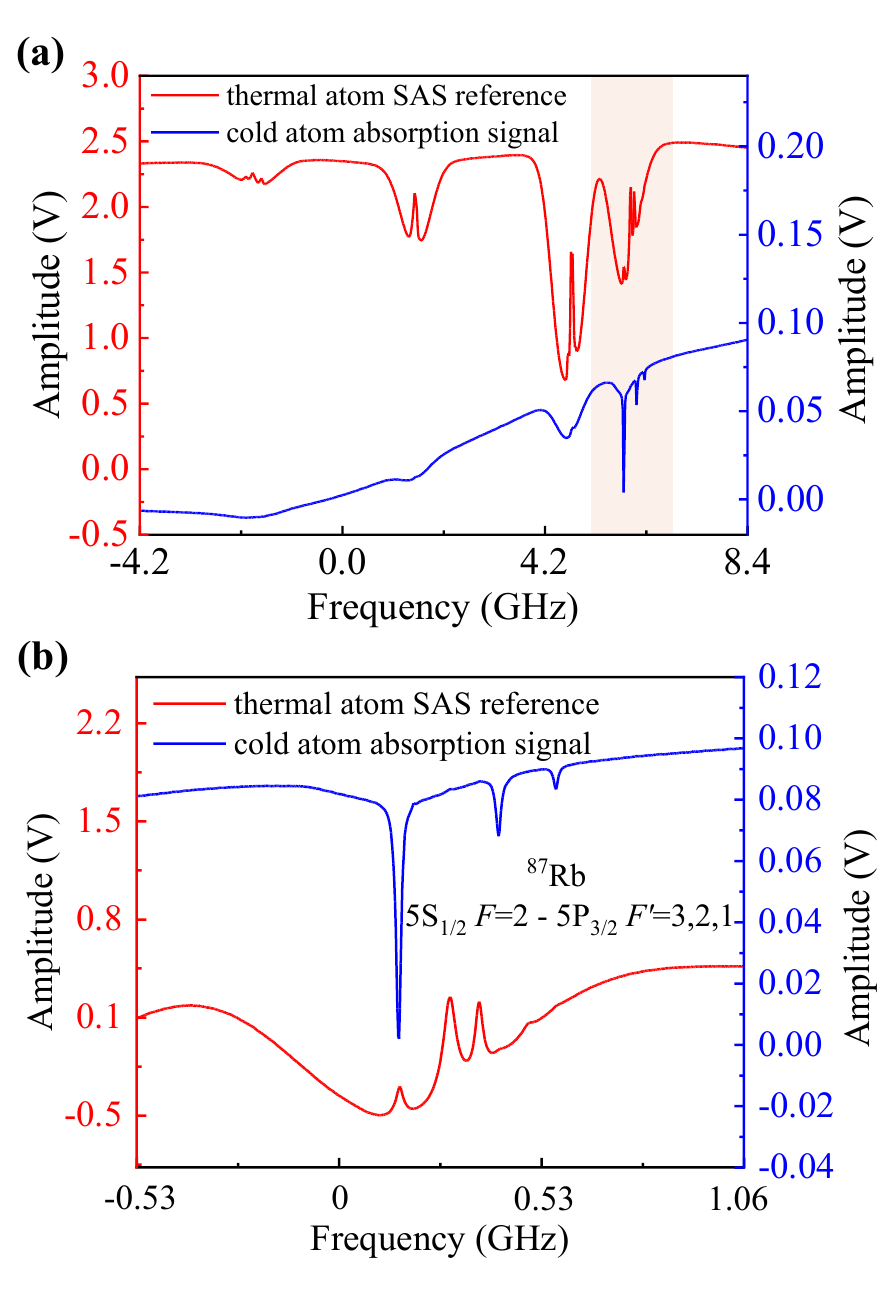}
	\caption{\label{figure-6}
		When a 780~nm laser is used as the cooling light, the absorption spectroscopy of cold $^{87}\mathrm{Rb}$ atoms for the 780~nm probe light is shown as the blue spectral line. The red spectral line is the saturated absorption spectroscopy of rubidium atoms used as a reference. (a) It contains all spectral lines of the 780~nm transitions of $^{85}\mathrm{Rb}$ and $^{87}\mathrm{Rb}$, with $^{87}\mathrm{Rb}$ atoms cooled to the $\mathrm{5S_{1/2}}(F=2)$ state. (b) For the expanded spectral lines, obvious cold atom absorption can be observed at the position corresponding to the $^{87}\mathrm{Rb}$ $\mathrm{5S_{1/2}}(F=2)\rightarrow\mathrm{5P_{3/2}}(F^{\prime}=1,2,3)$.
	}
\end{figure}

Experimentally, we also utilize another 780~nm laser as the cooling light to achieve cooling of $^{87}\mathrm{Rb}$ atoms. Since both the cooling light and the repump light operate in the 780~nm band, they are combined and then coupled into four 1$\times$2 (one-input, two-output) multimode optical fibers designed for the 780~nm wavelength. The light is directed into the vacuum atomic cell through optical access ports on the cell. The cooling light is tuned to a red detuning of approximately 15 MHz relative to the $^{87}\mathrm{Rb}$ $\mathrm{5S_{1/2}}(F=2)\rightarrow\mathrm{5P_{3/2}}(F^{\prime}=3)$ transition. The 780~nm repumping light still corresponded to the $\mathrm{5S_{1/2}}(F=1)\rightarrow\mathrm{5P_{3/2}}(F^{\prime}=2)$ transition. By precisely controlling the frequency and intensity of both the cooling and repump lights, a significantly improved cooling effect for $^{87}\mathrm{Rb}$ atoms is ultimately achieved. After the rubidium atoms are cooled, a 780~nm probe beam is similarly used to detect the cold atomic cloud. The resulting cold atom absorption spectroscopy is shown in Fig.~\ref{figure-6}. Due to the higher efficiency of 780~nm laser cooling, the number of cold atoms obtained is substantially increased. Consequently, three absorption peaks are visible, corresponding to the transitions $\mathrm{5S_{1/2}}(F=2)\rightarrow \mathrm{5P_{3/2}}(F^{\prime}=1,2,3)$, respectively.

\subsection{Optimization of the cooling light parameters}

\begin{figure}
	\includegraphics[width=0.9\linewidth]{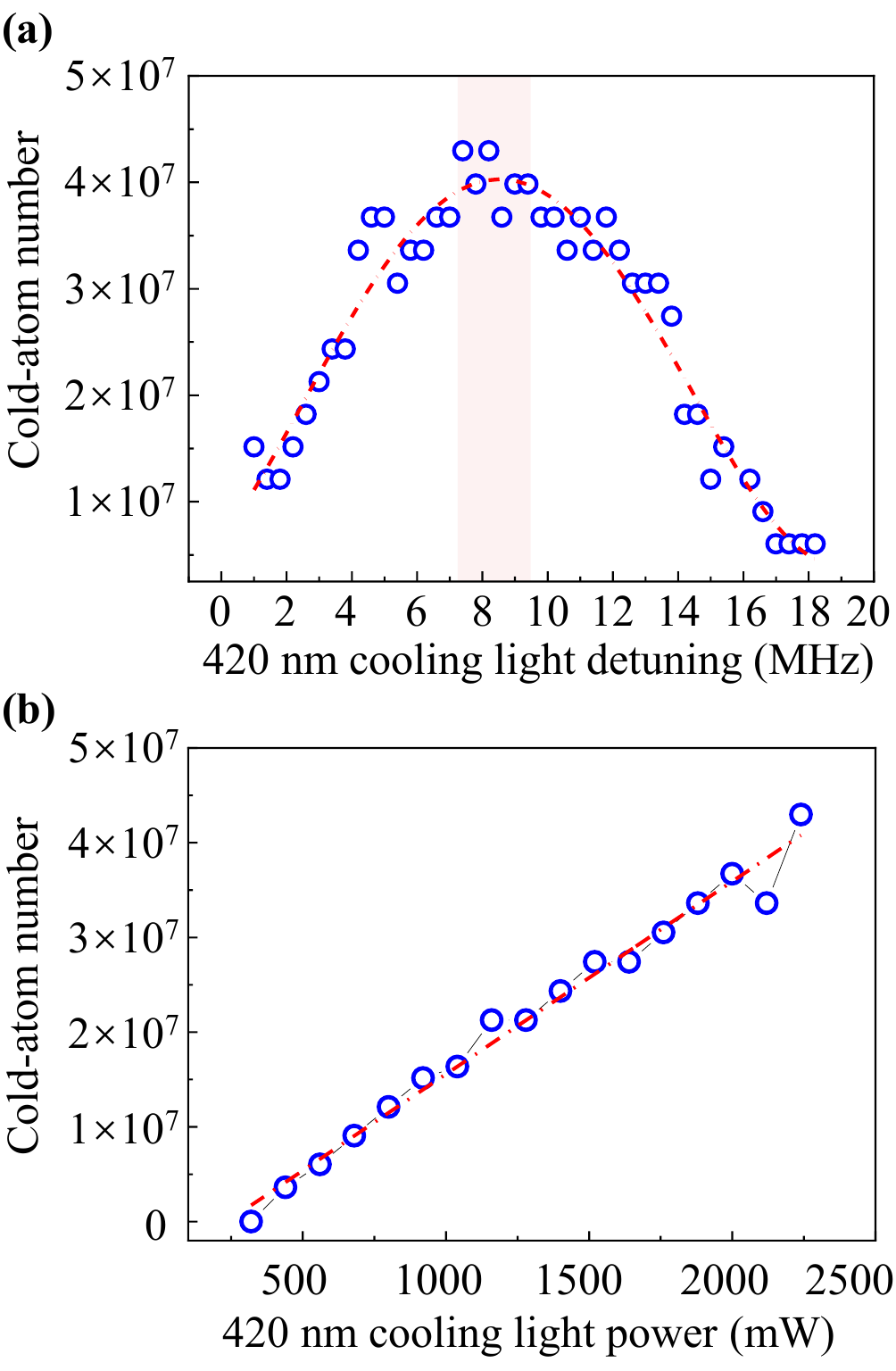}
	\caption{\label{figure-7}
		When using 420~nm blue light for laser cooling, (a) the absorption amplitude of cold $^{87}\mathrm{Rb}$ atoms for the 780~nm probe light varies with the detuning of the 420~nm cooling light. The cooling effect is optimal when the 420~nm cooling light is red-detuned by approximately 8~MHz relative to the $^{87}\mathrm{Rb}$ $\mathrm{5S_{1/2}}(F=2)\rightarrow\mathrm{6P_{3/2}}(F^{\prime}=3)$ transition. (b) the absorption amplitude cold $^{87}\mathrm{Rb}$ atoms for the 780~nm probe light as a function of the 420~nm cooling light power. In the cooling light power range of 0-2.5~W, no saturation phenomenon has been observed.
	}
\end{figure}

In cold atom physics experiments, the absorption characteristics of cold atoms for probe light serve as a key indicator for evaluating experimental performance. In the experiment of cooling $^{87}\mathrm{Rb}$ atoms with 420~nm blue light, through the analysis of the absorption spectroscopy of 780~nm probe light, not only can the quantitative evaluation of the number of cold atoms, but also an experimental basis can be provided for the optimization of cooling parameters. The red detuning of the 420~nm cooling light is a key parameter affecting the cooling efficacy. In the experiment, by adjusting the red detuning of the 420~nm cooling light relative to the $\mathrm{5S_{1/2}}(F=2)\rightarrow\mathrm{5P_{3/2}}(F^{\prime}=3)$ transition of $^{87}\mathrm{Rb}$ atoms (gradually increasing from 0~MHz to 18~MHz), the variations in the absorption amplitude of cold atoms for the 780~nm probe light are monitored (as shown in Fig.~\ref{figure-7}(a)). The experimental results show that the absorption amplitude presents a trend of first increasing and then decreasing with the increase of detuning, reaching its maximum when the detuning is 8~MHz. When the detuning is too small, the resonant interaction between the cooling light and atoms is too strong, leading to a significant recoil heating effect after atoms absorb photons, resulting in a small number of cold $^{87}\mathrm{Rb}$ atoms. When the detuning is around 8 MHz, the red detuning matches the Doppler shift of atomic thermal motion, the cooling force and damping force reach a balance, and the trapping efficiency of cold atoms is the highest. When the detuning is too large, the frequency of the cooling light deviates too far from the atomic resonant transition, the photon absorption probability decreases, and the cooling efficiency declines, causing the absorption amplitude to decrease.

Cooling light power is another important optimization parameter. In the experiment, the 3~W 420~nm laser undergoes frequency shifting, frequency locking, and multimode fiber coupling, resulting in a maximum delivered power of approximately 2.5~W injected into the vacuum atomic cell. By adjusting the cooling light power within the range of 0-2.5~W, the relationship between absorption amplitude and power is obtained (as shown in Fig.~\ref{figure-7}(b)). Different from the detuning experiment, the absorption amplitude shows a monotonic increasing trend with the increase of power, without any saturation phenomenon. This indicates that within the experimental power range, the cooling light power has not yet reached the saturation threshold for cold atom cooling. Increasing power enhances photon density per unit volume, thereby accelerating the photon scattering rate and simultaneously improving cooling efficiency, which increases the cold atom population. In the future, we plan to use a second 420~nm high-power laser as an additional cooling light to further boost the cooling power. However, further increasing the power may reduce the transition cross-section due to the light intensity saturation effect, or disrupt the cooling equilibrium due to excessive photon recoil force. Future experiments need to expand the power range to find the critical saturation point.

Cold-atom number, as a core parameter characterizing the absorption capability of cold atomic clouds, is directly related to the density $n$, the spatial distribution of the atomic cloud, and the transition cross-section $\sigma$. The density $n$ is defined as $n=-\mathrm{ln}(\frac{I_{\mathrm{trans}}}{I_{0}})/\sigma L$~\cite{hsiao2014cold}, where $I_{0}$ is the incident probe light intensity, $I_{\mathrm{trans}}$ is the transmitted light intensity, and where $L$ is the effective interaction length between light and atoms. 
When detecting cold atoms, the light intensity needs to be adjusted to the microwatt level to avoid excessive light intensity causing atom heating or absorption saturation. Meanwhile, during the detection process, attention must be paid to background light subtraction: additionally collect the dark field light intensity $I_{\mathrm{dark}}$ (only ambient light and detector dark current) when the probe light is turned off, and subtract it from $I_0$ and $I_{\mathrm{trans}}$. For a 780~nm probe light, $\sigma_{0}=\frac{3\lambda^{2}}{2\pi}$. Finally, it is derived that $n=-\frac{2\pi}{3\lambda^{2}}\mathrm{ln}(\frac{I_{\mathrm{trans}}-I_{\mathrm{dark}}}{I_{0}-I_{\mathrm{dark}}})$. Quantitative analysis via cold-atom number shows that when the 420~nm cooling light power is 2.5~W, the cold-atom density is about $1.4\times10^{11}\mathrm{m^{-3}}$ and the cold-atom number is approximately $4.4\times10^{7}$. Comparing the changes in absorption amplitude (cold-atom number) under different parameter combinations reveals that detuning optimization has a more significant impact on the number of cold atoms, while power enhancement can further increase atom cooling capacity on the basis of optimal detuning.

In addition, traditional 780~nm laser cooling is also performed. We investigate the variation of the absorption amplitude of cold atoms with the detuning and power of the cooling light. The cooling light power determines the intensity of the radiation pressure exerted by the light field on the atoms. When the power is too low, the photon scattering rate is insufficient, and atoms cannot be effectively decelerated and cooled. When the power is too high, it may induce additional heating, leading to atomic escape. Theoretically, for small detunings, although atoms can be cooled rapidly, the equilibrium temperature is relatively high; excessively large detunings reduce the scattering rate and prolong the cooling time. In addition, both small and excessively large detunings can affect the density of cold atoms.

\begin{figure}
	\includegraphics[width=0.85\linewidth]{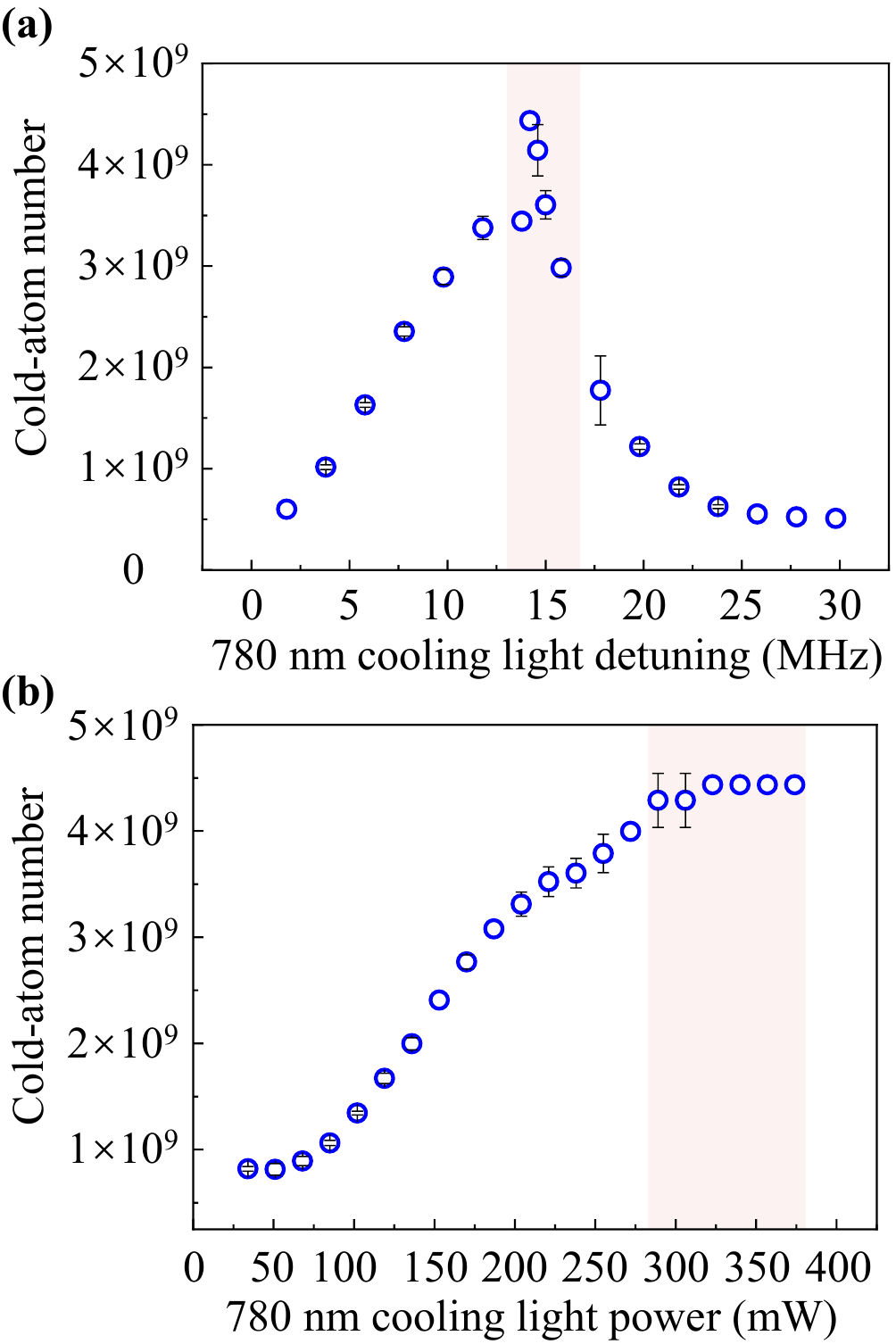}
	\caption{\label{figure-8}
		When using 780~nm laser for cooling, (a) the absorption amplitude of cold $^{87}\mathrm{Rb}$ atoms for the 780~nm probe light as a function of the detuning of the 780~nm cooling light. The cooling effect is optimal when the 780 nm cooling light is red-detuned by approximately 15~MHz relative to the $^{87}\mathrm{Rb}$ $\mathrm{5S_{1/2}}(F=2)\rightarrow\mathrm{5P_{3/2}}(F^{\prime}=3)$ transition. (b) The absorption amplitude of cold $^{87}\mathrm{Rb}$ atoms for the 780~nm probe light as a function of the power of the 780~nm cooling light. When the cooling light power increases to 300~mW, it reaches saturation.
	}
\end{figure}

In the experiment, the 780~nm cooling light power is fixed at 300~mW, and the detuning of the cooling light is adjusted from near resonance to large negative detuning, while measuring the cold-atom number. As the detuning increase, the cold-atom number gradually increase. When the detuning is 15~MHz (2.5 times the natural linewidth), the cold-atom number reached a maximum value of approximately $4.2\times10^{9}$, as shown in fig.~\ref{figure-8}(a). When the detuning exceeds 15~MHz, the cold-atom number gradually decrease until it reaches 0. With the cooling light detuning fixed at 15~MHz, the variation of the cold-atom number with the 780~nm cooling light power is obtained. Starting from a power of 0 and gradually increasing the cooling light power, the cold-atom number increase accordingly. At a cooling light power of 300~mW, the critical power value is reached, and the cold-atom number enters a saturated state, as shown in Fig.~\ref{figure-8}(b). When the power exceedes this value, the cold-atom number decrease due to the heating effect. Through quantitative analysis of the cold-atom number, it is found that when the 780~nm cooling light power is 300~mW and the optimal detuning is 15~MHz,the cold-atom density is about $1.3\times10^{13}\mathrm{m^{-3}}$ and the cold-atom number is about $4.2\times10^{9}$. 

The significant differences in the optimal detuning and saturation light power between 420~nm and 780~nm laser cooling may result from multiple factors. The momentum of a 420~nm photon is approximately 1.86 times that of a 780~nm photon. In laser cooling, the order of magnitude of the momentum change transferred to an atom in each absorption-spontaneous emission cycle is $\pm \hbar/\lambda$ (momentum is gained during absorption, and the average contribution of isotropic spontaneous emission is 0). Therefore, the change in atomic velocity (kinetic energy) per scattering event for 420~nm blue light is approximately 1.86 times that for 780~nm red light. Atoms need to be scattered more times to be decelerated to the same velocity (cooled to the same temperature) or require a higher scattering rate to overcome larger recoil disturbances. The optimal detuning for Doppler cooling needs to balance cooling efficiency (the damping coefficient is maximum at $\delta\approx-\Gamma/2$, but the trapping range is small) and the velocity trapping range ($\delta v\propto|\delta|\pm\Gamma/2$). The initial velocity distribution of thermal atoms is relatively broad, and the velocity trapping range ($\propto\Gamma$) of the 420~nm narrow-linewidth transition itself is narrow. Thus, it is necessary to increase $|\delta|$ to expand the trapping range ($\delta v \propto |\delta|$). Otherwise, high-speed atoms cannot be trapped. In addition, a large detuning can also suppress the total power and rate of recoil heating to a certain extent.

For the direct realization of cold rubidium atom using a 420 nm laser, a key advantage lies in its ability to directly serve the experimental platform of continuous cold-atom active optical clocks~\cite{chang2019stabilizing,zhang2024extremely}. Since the 420 nm laser assumes the dual role of cooling light and pumping light, it can directly excite rubidium atoms from the ground state ($\mathrm{5S_{1/2}}$) to the high excited state ($\mathrm{6P_{3/2}}$). Subsequently, the atoms transition to the intermediate state ($\mathrm{6S_{1/2}}$) via spontaneous emission. This process establishes a population inversion between the $\mathrm{6S_{1/2}}$ and $\mathrm{5P_{3/2}}$, $\mathrm{5P_{1/2}}$ energy levels, thereby providing the required gain medium for 1367~nm/1323~nm active optical clocks. This integrated cooling and pumping mechanism eliminates the reliance on additional pumping light sources, which is conducive to system integration and long-term operational stability. This scheme provides a more concise and robust technical route for continuous cold-atom active optical clocks.

\section{CONCLUSION}
This work surprisingly realizes the direct laser cooling of $^{87}\mathrm{Rb}$ atoms using 420~nm blue light for continuous cold-atom active optical clock, breaking through the traditional paradigm that relies on 780~nm laser cooling (corresponding to the first excited state $\mathrm{5P_{3/2}}$). First, to the best of our knowledge, this is the first time that a 420~nm laser has been directly utilized as cooling light to achieve cold $^{87}\mathrm{Rb}$ atoms. Second, as an important step towards our subsequent cold-atom active optical clock. This system simultaneously realizes population inversion in the gain medium atoms between the $\mathrm{6S_{1/2}}$ and $\mathrm{5P_{3/2}}$, $\mathrm{5P_{1/2}}$ energy levels during the cooling process, corresponding to the 1367~nm and 1323~nm active optical clock lasers.
The core innovation lies in performing cooling operations directly based on the transition of rubidium atoms to the high excited state ($\mathrm{6P_{3/2}}$). Through a carefully designed 420~nm diffuse laser cooling scheme, we experimentally obtained a cold $^{87}\mathrm{Rb}$ atomic cloud with a length of up to one meter, which lays an important foundation for large-scale quantum coherent manipulation. The measurement of the absorption spectroscopy of the cold atomic cloud in the experiment directly confirms the cooling effect.  At the current stage, 420~nm blue light diffuse laser cooling can reach a cold-atom number of $4.4\times10^{7}$. 

The most significant implication of this work is the first direct and effective cooling of ground-state rubidium atoms using 420~nm blue light, while simultaneously exciting them to the 6P state. This innovative method circumvents the traditional approach of utilizing the first excited state of rubidium for cooling or the sequential path of 780~nm cooling followed by 420~nm cooling. It thereby opens a new technical route for the efficient and direct preparation and coherent manipulation of Rydberg atoms. It is also possible to achieve simultaneous cooling and pumping in cold-atom active optical clock system. This cooling technique directly connects the ground state with highly excited states and promises impactful applications in multiple frontier fields: ultracold quantum gases and novel quantum states, quantum information processing, cold-atom active optical clock, cold-atom quantum frequency standard, and the study of many-body quantum dynamics.

\vspace{1cm}
\textbf{Data availability}
Data underlying the results of this study are available from the authors upon request.

\textbf{Acknowledgments}
This work was funded by National Natural Science Foundation of China (Grant Nos. 624B2010, 62405007), the Innovation Program for Quantum Science and Technology (2021ZD0303200).

\textbf{Author contributions}
J.Chen conceived the idea to directly use a 420~nm laser as cooling light to cool rubidium atoms. J.Zhang and X.Gao performed the experiment. J.Zhang wrote the manuscript. Z.Xiao helped with theoretical analysis. X.Guan, R.Chen and M.Han helped with laser cooling experiment. T.Shi and J.Chen provided revisions.

\textbf{Competing interests}
The authors declare no competing interests.


\bibliography{blue-laser}

\end{document}